\documentclass[prd,reprint,showpacs,showkeys,twocolumn]{revtex4}
\usepackage{amsfonts}
\usepackage{amssymb}
\usepackage{amsmath}
\usepackage{graphicx}
\usepackage{float}
\usepackage[font={footnotesize,it}]{caption}
\usepackage{tikz}
\usepackage{tikz-3dplot}
\usepackage[colorlinks=true,
            linkcolor=purple,
            urlcolor=purple,
            citecolor=blue]{hyperref}

\begin{document}

\title{Thin-Shells and Thin-Shell Wormholes in New Massive Gravity}
\author{S. Danial Forghani}
\email{danial.forghani@emu.edu.tr}
\author{S. Habib Mazharimousavi}
\email{habib.mazhari@emu.edu.tr}
\author{M. Halilsoy}
\email{mustafa.halilsoy@emu.edu.tr}
\date{\today }
\affiliation{Department of Physics, Faculty of Arts and Sciences, Eastern Mediterranean
University, Famagusta, North Cyprus via Mersin 10, Turkey}

\begin{abstract}
Within 2+1-dimensional cosmological new massive gravity, we consider
thin-shell and thin-shell wormhole construction. For this, we introduce
first, the junction conditions apt for the fourth order terms in the action
of the theory. Then, by employing some specific static solutions in new
massive gravity, we study the characteristics of associated thin-shells and
thin-shell wormholes. Our finding suggests that, firstly, there cannot exist
any thin-shells regarding our chosen solutions of cosmological new massive
gravity, and secondly, the constructed thin-shell wormhole does not need to
be symmetric. More importantly, the thin-shell wormhole, if ever forms,
possesses null energy density and null angular pressure on its throat which
preferable to their negative-valued counterparts.
\end{abstract}

\maketitle

\section{Introduction}

Spherically symmetric thin-shells (TSs) have been the subject of many
interesting studies in the literature. For instance, long ago, in 1965,
Arnowitt, Deser and Misner showed that the self-energy of a charged dust
shell is finite \cite{Arnowitt}. In 1973 Boulware in his short paper \cite%
{Boulware} studied a charged TS whose inside and outside spacetimes are flat
Minkowski and Reissner-Nordstr\"{o}m, respectively. In there, using the
Israel formalism, the dynamics of the shell was studied and it was shown
that if the matter energy density of the shell is negative then the shell
may collapse to form a naked singularity. A general spherical shell,
however, was studied by Lake in 1978 \cite{Lake}. Lake showed that, imposing
positive definite total proper mass for the shell implies the impossibility
of the merge of the inner and black-hole horizons. In \cite{Heusler},
Heusler \textit{et al.} calculated the self-energy of a charged TS with flat
spacetime inside and Reissner-Nordstr\"{o}m outside. While their work for a
dust shell gives the results of \cite{Arnowitt}, some of their findings were
already found in other papers \cite{Kuchar, Chase}. Later on, spherically
symmetric TS in Brans-Dicke theory of gravity was considered by Letelier and
Wang in \cite{Wang}. Furthermore, the linear stability analysis of
spherically symmetric timelike shells and bubbles attracted attentions \cite%
{Simeone}. For instance, in \cite{Brady}, the stability of a TS surrounding
a Schwarzschild black hole was studied. It was shown that such a TS may be
stable against a radial perturbation if its radius is larger than the radius
of the circular photon orbit. Finally we would like to mention the
collapsing of the higher-dimensional spherically symmetric TS with rotation
which has been studied recently \cite{Delsate}. Such kind of study in $3+1-$%
dimensions with different symmetry, prior to \cite{Delsate}, had already
been studied in \cite{Nolan, Mena, Prisco}.

Wormholes appeared in general relativity as special solutions of the
Einstein's field equations. These bizarre structures can be visualized as
tunnels connecting remote points within a single spacetime or points of two
distinct spacetimes. The main problem of wormholes is that they are
supported by exotic matter; a kind of matter which does not satisfy known
energy conditions. In 1988, Morris and Thorne published a paper and
discussed traversability of wormholes \cite{Morris1}. The paper soon came to
focus of researchers and caused a new wave of studies over structural
characteristics of wormholes. In 1989, in one of these attempts, Visser
developed a method nowadays known as the cut-and-paste procedure, with which
a new class of traversable wormholes came into existence \cite{Visser1}. The
idea of these so-called thin-shell wormholes (TSWs) gained much popularity
since its original construction. The principal aim was how to tackle with
the exotic matter that necessarily gave life to such an object. Visser's
recipe was to confine the exotic matter to a very narrow surface, i.e. the
thin-shell, so that its existence can be justified in some way. Another
advantage of TSWs is that they can be constructed by a vast variety of
spacetimes, as there are already many TSWs appearing in the literature \cite%
{Richarte1,Varela1,Sharif1,Sharif2,Montelongo1,Dias1,Eiroa1,Thibeault1,Lobo1,Eiroa2,Eiroa3,Ishak1,Poisson1}%
. However, almost all TSWs suffer from two serious drawbacks: the exotic
matter as source, and fragile/limited and/or non-physical stability
characteristics. This point is precisely the reason that the challenge of
finding stable TSWs is still going on, hopefully, with no exotic matter on
them. In this regard, we make our search of TSWs in a special, modified
massive theory of gravity. (For more details on wormholes and TSWs the
reader is advised to consult \cite{Visser2}\ and references therein.)

The mass of the quantum gravity's fundamental particle, the graviton, has
been one of the most disputable subjects of modern physics. In massive
theories of gravity, the spin-2 graviton is destined to move inside the
local light cone, not on it. To point out the importance of massive theories
of gravity let us draw a rough analogy with the Standard Model (SM) of
particle physics by recalling that neutrino shared a similar history. In the
SM, neutrino was also thought to be massless for a long time. With the
advent of experimental neutrino physics, the picture has changed: neutrino
has a very small but non-zero mass \cite{Mertens1}. This amounts to changing
much of the rules in the SM, leading even to revise certain proportion of
the textbooks. In analogy, with massive gravity a certain revision in the
fundamental physics of gravitational waves is expected.

To prove the mass of a graviton, however, is more challenging than the
neutrino. In the gravity side, even at a classical level, we had to wait
until very recently when LIGO and Virgo detected gravitational waves from
the merger of massive black holes in distant past \cite{LIGO1,Virgo1}.
Assuming that the gravity waves are massive, their speed in vacuum will be
less than the speed of light and this will naturally cause a delay in their
arrival to the Earth. Nonetheless, an observation by LIGO and Virgo in 2017,
put a constraint on this time delay, limiting the difference between the
speed of gravitons and the speed of photons to the interval $\left( -3\times
10^{-15},+7\times 10^{-16}\right) $-times the speed of light \cite{Virgo1}.
This means, that there still is an uncertainty in the mass of gravitons, and
until the time the precision of our measurements lets us decisively confirm
or reject the existance of massive gravitons, theoretical physics will keep
contributing to the concept. See \cite{Hinterbichler1,de Rham1} as review
studies, and references therein for more details on massive gravity.

In literature, it was Fierz and Pauli who added a mass dependent term to the
gravity action for the first time in 1939 \cite{Fierz1}. Since then, there
has been so many attempts to establish a consistent quantum gravity theory,
especially in 3+1 dimensions. However, most of these theories suffer from a
common disadvantage: absence of renormalizable theory. While renormalization
is a big problem in 3+1 dimensions, things are different in 2+1. From simple
power counting method of field theory, the lower dimension expectedly has
natural advantages over the higher dimension \cite{Gavela1}. Since
2+1-dimensional Weyl curvature vanishes identically, it is well-known that
there are no pure gravitational degrees of freedom. For this reason, in
order to create a theory, source must be supplied in the lower dimension to
make a gravitationally feasible theory. This is done by different methods,
among which, two received more endorsements.

In an attempt to construct such a theory in 2+1 dimensions, Deser \textit{et
al.} established the theory of Topological Massive Gravity (TMG) in 1982 by
adding a Chern-Simons term to the 3-dimensional Einstein-Hilbert (EH) action 
\cite{Deser1,Deser2}. In 2009, Bergshoeff \textit{et al.} suggested a
different theory which later was known as New Massive Gravity (NMG) \cite%
{Bergshoeff1,Bergshoeff2}. This theory, which at the linearized level is the
2+1 dimensional equivalent of Pauli-Fierz theory, has the advantage over TMG
that preserves parity symmetry \cite{Bergshoeff1,Bergshoeff2,Bergshoeff4}.
Furthermore, it is shown in \cite{Oda1} and \cite{Nakasone1}\ that the
theory is unitary in tree level and renormalizable. de Rham \textit{et al.} 
\cite{de Rham2} go further to indicate that NMG is unitary even beyond the
tree level. In \cite{Accioly1} the authors prove that NMG at tree level is
actually the only 3-dimensional unitary system which can be constructed by
adding quadratic curvature terms to EH action. Add to all these remarkable
characteristics, its invariance under general coordinate transformations is
also manifest.\ NMG, therefore, gained much attentions right after its
introduction, for being a promising candidate for a renormalizable theory of
quantum gravity. The theory also exhibits features like gravitational time
dilation and time delay which the usual 3-dimensional general relativity is
not subject to \cite{Accioly1,Accioly2}. Soon after the first publication,
it was shown that it admits exact black hole solutions. In \cite{Clement1}
warped AdS$_{3}$ black holes and AdS$_{2}\times $S$^{1}$ solutions, in \cite%
{Bergshoeff2} Ba\~{n}ados-Teitelboim-Zanelli (BTZ) \cite{Banados1}, new type
black holes, and warped dS$_{3}$ and dS$_{2}\times $S$^{1}$ solutions, in 
\cite{Clement2} extreme BTZ and a family of massive `log' black holes, and
in \cite{Ayon-Beato2}\ and \cite{Clement2}\ AdS waves are discussed. Another
important contribution is made by Oliva \textit{et al.} who investigated
exact black hole and non-black hole solutions of a special case with
negative, positive and vanishing cosmological constant in \cite{Oliva1}.
Moreover, the Lifshitz metrics have been shown to be solutions of NMG for
generic values of the dynamical exponent $z$ (with an exact - asymptotically
Lifshitz - black hole solution at $z=3$) \cite{Ayon-Beato1}. Ahmedov and
Aliev in a series of exquisite papers \cite{Ahmedov1,Ahmedov2,Ahmedov3}
discuss algebraic type D and type N solutions of NMG by employing a
first-order differential Dirac-type operator. They indicate that the NMG
field equations can be considered as square of TMG field equations, and
accordingly, argue the possibility of mapping all types D and N solutions of
TMG into NMG. Besides, they find new types D and N solutions in NMG with no
counterparts in TMG. The stability of BTZ black holes in NMG\ are
classically studied in \cite{Myung1}. Some of these aforementioned solutions
will be considered here in this study.

Among all these, AdS$_{3}$ solutions are of greater importance for an
obvious reason: Where there is a quantum-gravity consistent theory along
with AdS solutions, there exists the AdS/CFT correspondence \cite{Maldacena1}%
. However, it was shown that on the boundary of the dual CFT, the unitarity
of the AdS vacuum connote a negative central charge \cite{Bergshoeff2,Liu1}.
Also, for logarithmic CFT correspondence (AdS$_{3}$/LCFT$_{2}$) see \cite%
{Grumiller1}. Setare and Kamali in \cite{Setare2} show that there is a
perfect agreement between their results using 2-dimensional Galilean
conformal algebra on the boundary of NMG with the Bekenstein-Hawking entropy
(in nonrelativistic limit) for warped AdS$_{3}$ and contracted BTZ black
hole solutions of NMG. Phase transitions between BTZ black hole solutions
and thermal solitons within NMG are studied in \cite{Eune1,Myung2,Zhang1}.
It is also worth-mentioning that later, the new type black holes initially
appeared in \cite{Bergshoeff2}, came to the attention of Kwon \textit{et al.,%
} who obtained their quasi-normal modes \cite{Kwon1}, and Gecim and Sucu,
who studied the properties of relativistic spin-1/2 and spin-0 particles in
this background \cite{Gecim1}.

NMG has also been generalized to 4th \cite{Bergshoeff3} and higher arbitrary
dimensions \cite{Dalmazi2}. Although, it is summed up in \cite{Nakasone1}
that the higher dimensional generalizations are not unitary at the tree
level.

Furthermore, there have been attempts to extend NMG. Among these, we point
out the novel works by G\"{u}ll\"{u} \textit{et al.} \cite{Gullu1,Gullu2},
which extend NMG to a 3-dimensional Born-Infeld theory of gravity (BI-NMG).
There, the authors discuss that the cubic order extension of their augmented
action duplicates the deformation of the NMG gained from AdS/CFT
correspondence. Exact black hole solutions of BI-NMG are discussed in \cite%
{Ghodsi1}, where properties such as mass, angular momentum, entropy and CFT
dual central charges of the solutions are also determined. Extensions to
higher curvature theories (R$^{3}$-NMG) and their exact solutions are
considered in \cite{Sinha1,Nam1,Anastasiou1,Setare1}. In \cite{Dalmazi1}
even higher derivative kinetic terms are discussed. Algebraic type N
spacetime solutions to BI-NMG and their higher order curvature corrections
of NMG are studied in \cite{Ahmedov4}. An extension of the theory by scalar
matter with Higgs-like self-interaction is investigated in \cite{Louzada1}
with exact asymptotically dS$_{3}$ solutions which qualifies as an eternally
accelerated non-singular bounce-like 3D Universe. Another extension by
scalar matter is discussed in \cite{Camara1}, where the authors study a
family of flat static domain walls as solutions. In \cite{Ghodsi2}, the NMG
action is coupled to Maxwell's electromagnetic and Chern--Simons actions to
give rise to charged black holes in both warped AdS$_{3}$ and log forms.
Generalized Massive Gravity (GMG), whose action contains quadratic terms of
both TMG and NMG along with coupling constants, and all its homogenous
solutions are studied in \cite{Bakas1}. Exploiting the NMG action, a new
bi-gravity model is constructed in 3 dimensions in \cite{Akhavan1}. Finally,
a novel work by Dereli and Yeti\c{s}mi\c{s}o\u{g}lu suggests a model (new
improved massive gravity (NIMG))\ which includes TMG, NMG and minimal
massive gravity (MMG) as subclasses of the theory \cite{Dereli1,Dereli2}.

In this paper we particularly consider the cosmological new massive gravity
(CNMG) in 2+1-dimensions \cite{Bergshoeff2}\ and construct TSs and TSWs in
such a theory.

Our findings are interesting from physics point of view. On one hand, we
find that the propounded metrics admit no TS. This TS-nonexistence is
particularly important when it holds for AdS metrics. On the other hand, we
find TSWs within CNMG which can be stable, but come to exist only when are
asymmetric, in the sense that the bulk spacetimes on the two sides are
different in geometry and nature. These rather new type of TSWs are called
asymmetric TSWs (ATSWs). As spherical and cylindrical ATSWs in general
relativity, these are explicitly considered in \cite%
{Forghani1,Forghani2,Forghani3}, while spherical ATSWs in $F(R)$ gravity are
studied implicitly in \cite{Eiroa4}. The tidbit is that for the cases we are
studying here, such ATSWs do not need matter (neither ordinary nor exotic)
as support, and hence, provide a smooth passage from one universe to the
other. Stated otherwise, our TS with zero energy-momentum acts as vacuum as
long as it is not perturbed. Once perturbed the perturbation energy will
accumulate a non-zero energy-momentum on the TS.

To achieve this aim, we choose a class of static solutions in CNMG and
introduce the necessary junction conditions (JCs) apt for a higher order
theory. This amounts to revision of Darmois-Israel JCs (DI-JCs) that were
designed for Einstein's general relativity \cite{Israel1}. The qualified JCs
in quadratic gravity are mentioned in \cite{Deruelle1}, and later in 2016,
revised in \cite{Reina1}. It is worth mentioning that Eiroa \textit{et al.}
successfully applied the latter JCs to establish TSWs with a double layer 
\cite{Eiroa4}, pure double-layer bubbles \cite{Eiroa5}, and spherical TS 
\cite{Eiroa6} in $F\left( R\right) $ theory of gravity.

The paper is organized as follows. In section $II$ we briefly introduce the
CNMG theory and a class of static solutions of the theory. Construction of
TS and TSWs are generally discussed in section $III$. Section $IV$ is
devoted to introducing the proper JCs in CNMG, which are applied to
legitimize the details of the TS and TSW's constructions in subsections.
Finally, section $V$ completes the paper with our conclusion.

\section{Cosmological New Massive Gravity Solutions}

The NMG theory is based on the $2+1$-dimensional cosmological new massive
gravity (CNMG) action \cite{Bergshoeff2}%
\begin{multline}
I_{\text{CNMG}}=\frac{1}{2\kappa }\int d^{3}x\sqrt{-g} \\
\left( \varsigma R+\frac{1}{m^{2}}\left( R_{\mu \nu }R^{\mu \nu }-\frac{3}{8}%
R^{2}\right) -2\lambda m^{2}\right) ,
\end{multline}%
in which $\kappa ^{-1}$ is the three dimensional reduced Planck mass, $m$ is
the mass of the graviton, $R_{\mu \nu }$ and $R$ define the Ricci tensor and
the Ricci scalar, respectively, and $\lambda $\ is a dimensionless
cosmological parameter. The factor $\varsigma $ in Einstein-Hilbert term is
merely a convention dependent factor which takes on either $1$ or $-1$.

In this piece of work we consider the solutions of the theory which can be
cast into the generic circularly symmetric\ form%
\begin{equation}
ds^{2}=-f\left( r\right) dt^{2}+\frac{1}{f\left( r\right) }%
dr^{2}+H^{2}\left( r\right) d\theta ^{2},
\end{equation}%
where%
\begin{equation}
f\left( r\right) =c_{0}+c_{1}r+\frac{1}{2}c_{2}r^{2}
\end{equation}%
and $H\left( r\right) $ are functions of the radial coordinate $r$. Herein, $%
c_{0}$\ and $c_{1}$\ are integration constants to be interpreted as the mass
parameter and gravitational hair of the given spacetime, respectively. The
cosmological-like parameter $c_{2}$ can be reparametrized by the
cosmological parameter $\lambda $\ as 
\begin{equation}
c_{2}=4m^{2}\left( \varsigma \pm \sqrt{1+\lambda }\right) .
\end{equation}%
Setting $H\left( r\right) =r$ comprises BTZ, warped (A)dS$_{3}$ and new type
black hole solutions, while setting $H\left( r\right) =1$ recovers non-black
hole (A)dS$_{2}\times $S$^{1}$ solutions. These are explained below in more
details. Note that, in general, for $\lambda <-1$ there is no solution, and
for $\lambda =0$ the metric represents a flat spacetime.

$H\left( r\right) =r$:

a) For $\lambda >-1$ we must have $c_{1}=0$. These solutions for $\varsigma
=1$\ are locally isometric to AdS$_{3}$ and for $\varsigma =-1$ are locally
isometric to dS$_{3}$. In the special case of $c_{0}=1$ one recovers (A)dS$%
_{3}$ vacua. Furthermore, in AdS$_{3}$ case $c_{0}<1$ admits static BTZ
black holes with mass parameter $\mu =-c_{0}$. Also, for $\lambda =0$\ the
solution is trivially flat.

b) For $\lambda =-1$ the solutions are called new type black holes. These
special vacuum solutions for $\varsigma =1$ ($c_{2}>0$)\ exhibit
asymptotically AdS$_{3}$ unique vacua, while they are asymptotically dS$_{3}$
for $\varsigma =-1$ ($c_{2}<0$). In this case, the metric function $f\left(
r\right) $ , provided $c_{1}^{2}-8\varsigma m^{2}c_{0}\geq 0$, has a real
double root at%
\begin{equation}
r_{1,2}=\frac{1}{4m^{2}}\left( -\varsigma c_{1}\pm \sqrt{c_{1}^{2}-8%
\varsigma m^{2}c_{0}}\right) .
\end{equation}%
In AdS$_{3}$ case, when $r_{1}>0$ we have an asymptotically AdS black hole
with its horizon at $r=r_{1}$. There will also be an inner horizon at $%
r=r_{2}$ in case $r_{2}>0$. For dS$_{3}$, on the other hand, there
potentially exist two horizons. When $r_{1}>0$, the surface $r=r_{1}$ is
similar to the cosmological horizon of dS spacetime. If also $r_{2}>0$, we
will have a black hole with an event horizon at $r=r_{2}$. In this case, the
occurrence of double roots implies that in between the roots, $r_{2}<r<r_{1}$%
, the static spacetime remains static, whereas for $r<r_{2}$ and $r>r_{1}$\
becomes dynamic with $t\longleftrightarrow r$.

$H\left( r\right) =1$:

For $\lambda =-1$ the metric is Kaluza-Klein (KK) type vacuum solution, i.e.
locally isometric to AdS$_{2}\times $S$^{1}$ when $\varsigma =1$ ($c_{2}>0$) 
\cite{Clement1}, and to dS$_{2}\times $S$^{1}$ when $\varsigma =-1$ ($%
c_{2}<0 $) \cite{Bergshoeff2}.

Let us note that $c_{0}$ and $c_{1}$ are parameters of the solution, while $%
c_{2}$ is related to the essential parameters of the theory, $m$\ and $%
\lambda $.

\section{Thin-Shell and Thin-Shell Wormhole Construction}

Since the construction procedure of TSWs has been given extensively and
repeatedly in the literature \cite{Poisson1}, we shall make our presentation
in this section very brief.

In general, construction of TS and TSW have similarities and differences. To
construct a TS, consider two distinct Lorentzian spacetimes denoted by $%
\left( \Sigma ,g\right) ^{-}=\{x^{\mu }|r\leq a\}$ as the inner and $\left(
\Sigma ,g\right) ^{+}=\{x^{\mu }|r\geq a\}$ as the outer spacetimes, which
are distinguished by their common hypersurface $\partial \Sigma =\{x^{\mu
}|r=a\}$ such that $\partial \Sigma \subset $ $\left( \Sigma ,g\right) ^{\pm
}$. These two necessarily non-symmetric spacetimes represent smooth
manifolds and contain no singularities, event horizons or kinks. The
coordinates of these two spacetimes are not necessarily the same and will be
denoted by $x_{\pm }^{\mu }$. The line element on the hypersurface $\partial
\Sigma $\ (the TS) is given by%
\begin{equation}
ds^{2}=h_{ij}^{\pm }d\xi ^{i}d\xi ^{j},
\end{equation}%
where $\xi ^{i}$ are the local coordinates on the shell and $h_{ij}^{\pm }=%
\frac{\partial x_{\pm }^{\mu }}{\partial \xi ^{i}}\frac{\partial x_{\pm
}^{\nu }}{\partial \xi ^{j}}g_{\mu \nu }^{\pm }$ is the localized metric of $%
\partial \Sigma $. In the next section we will discuss that in fact $%
h_{ij}^{-}=h_{ij}^{+}$ must hold on the shell. The unit normal to the
surface is also given by $n_{\mu }^{\pm }\frac{\partial x_{\pm }^{\mu }}{%
\partial \xi ^{i}}=0$; $n_{\mu }^{\pm }n_{\pm }^{\mu }=1$.

In case of a TSW, we cut a region of each spacetime, such that $\left(
\Sigma ,g\right) ^{\pm }=\{x^{\mu }|r\geq a>r_{e}\}$, in which $r_{e}$ is
any existed event horizon. Then we glue the two regions at their common
hypersurface $\partial \Sigma \subset $ $\left( \Sigma ,g\right) ^{\pm }$
which usually is referred to as the throat of the TSW. Note that in most
cases the two separated regions $\left( \Sigma ,g\right) ^{\pm }$ are copies
of each other but this is not compulsory \cite{Forghani1}. Similar to the TS
case, one finds the metric on the throat unique as approaching it, no matter
from which side. The line element and the normal can be introduced in the
same way as for a TS. The key difference between a TS and a TSW is that in a
TS only one of the normals is chosen (for instance the one going into $%
\left( \Sigma ,g\right) ^{+}$ and out of $\left( \Sigma ,g\right) ^{-}$),
while for a TSW both normals are considered independently. Therefore, while
passing across the shell, the normal vector is continuous in one case (TS)
and discontinuous in the other (TSW). In TSW, this distinction between
normals is transmitted through all the extrinsic properties of the throat,
for the normals play parts in them, by definition. Hence, one must be
careful to hold the ($\pm $) signs of the normals for a TSW, while they can
be dropped casually for the case of a TS. For the intrinsic properties (such
as Riemann or Ricci tensors and Ricci scalar), of course, this is not the
case.

The extrinsic curvature tensor of the TS(W) is given by%
\begin{equation}
K_{ij}^{\pm }=-n_{\lambda }^{\pm }\left( \frac{\partial ^{2}x_{\pm
}^{\lambda }}{\partial \xi ^{i}\partial \xi ^{j}}+\Gamma _{\alpha \beta
}^{\lambda \pm }\frac{\partial x_{\pm }^{\alpha }}{\partial \xi ^{i}}\frac{%
\partial x_{\pm }^{\beta }}{\partial \xi ^{j}}\right) ,
\end{equation}%
where $\Gamma _{\alpha \beta }^{\lambda \pm }$ are the Christoffel symbols
of the bulk spacetimes, compatible with $g_{\alpha \beta }^{\pm }$.

In the following section we shall introduce the proper JCs for the static
solutions within NMG framework.

\section{The Junction Conditions}

In this section, we introduce two distinct sets of JCs, independently
derived in \cite{Deruelle1}\ and\ \cite{Reina1} qualified for quadratic
theories of gravity in arbitrary dimensions. In the latter, Reina, Senovilla
and Vera (RSV) take advantage of the standard distributional analysis, while
in the former, Deruelle, Sasaki and Sendouda (DSS) simplify the problem by
using Gaussian coordinates at the joint hypersurface. In \cite{Reina1} RSV
argue that using Gaussian coordinates often causes ignoring some important
subtleties, and therefore, the reliability extend of the method is ambiguous
(specially, when it comes to double layers). Nevertheless, since we are not
considering double layers, for the sake of curiosity we will apply both
methods, independently, and count similarities and differences, if there is
any. In what follows we particularly concentrate on timelike hypersurfaces
which make more physical sense.

\subsection{THE RSV Junction Conditions}

According to \cite{Reina1}, a general quadratic Lagrangian density in $n+1$
dimensions has the form%
\begin{multline}
\mathcal{L}_{\text{RSV}}=\sqrt{-g}\times \\
\left( R+a_{1}R^{2}+a_{2}R_{\mu \nu }R^{\mu \nu }+a_{3}R_{\alpha \beta \mu
\nu }R^{\alpha \beta \mu \nu }-2\Lambda \right) ,
\end{multline}%
where $a_{n}$ and $\Lambda $ are constants with physical dimensions of $%
\left[ L^{2}\right] $. A quick comparison with the CNMG action given in Eq.
(1), reveals that $a_{1}=\frac{-3\varsigma }{8m^{2}}$, $a_{2}=\frac{%
\varsigma }{m^{2}}$ and $a_{3}=0$. Similar to the normal ID-JCs in general
relativity, the diffeomorphism\ of the two spacetimes to be joined at the
junction requires the first fundamental form to be continuous at their
common hypersurface. This is the first JC and guarantees the identification
of a global metric in the sense of distributions. To avoid non-physical
distributional terms, in the case either $a_{2}$ or $a_{3}$ is nonzero, the
second JCs are identified as the continuity of the second fundamental form
at the junction. There are also other JCs which basically insure that the
other fundamental generalized functions are well-defined, as well. RSV also
introduce the parameters $\kappa _{1}=2a_{1}+a_{2}/2$ and $\kappa
_{2}=2a_{3}+a_{2}/2$ and classify their JCs in the case of a
\textquotedblleft \textit{TS without double layer}\textquotedblright\ based
on the values of $\kappa _{2}$\ and $n\kappa _{1}+\kappa _{2}$. With regard
to the coefficients $a_{1}$, $a_{2}$ and $a_{3}$, the proper JCs for
2+1-dimensional NMG are the ones for the case \textquotedblleft $\kappa
_{2}\neq 0$\textit{\ and }$n\kappa _{1}+\kappa _{2}=0$\textquotedblright .\
This is very interesting in the sense that even RSV would not think of this
case as a serious one:

\begin{quote}
\textquotedblleft \textit{Nevertheless, the relevance of this exceptional
case is probably marginal, as the coupling constants satisfy a dimensionally
dependent condition.}\textquotedblright
\end{quote}

\subsubsection{TS construction with RSV-JCs}

Accordingly, for investigating a TS, the proper JCs for $n=2$ will be 
\begin{subequations}
\begin{equation}
\left[ g_{\alpha \beta }\right] _{-}^{+}=0;
\end{equation}%
\begin{equation}
\left[ K_{\alpha \beta }\right] _{-}^{+}=0;
\end{equation}%
\begin{multline}
\left[ R_{\alpha \beta \mu \nu }\right] _{-}^{+}=\frac{\left[ R\right]
_{-}^{+}}{4}\times \\
\left( n_{\alpha }n_{\mu }h_{\beta \nu }-n_{\beta }n_{\mu }h_{\alpha \nu
}-n_{\alpha }n_{\nu }h_{\beta \mu }-n_{\beta }n_{\nu }h_{\alpha \mu }\right)
;
\end{multline}%
\begin{equation}
\left[ R_{\alpha \beta }\right] _{-}^{+}=\frac{\left[ R\right] _{-}^{+}}{2}%
\left( \frac{1}{2}h_{\alpha \beta }+n_{\alpha }n_{\beta }\right) ;
\end{equation}%
\begin{multline}
\left[ \nabla _{\mu }R_{\alpha \beta }\right] _{-}^{+}=n^{\nu }\left[ \nabla
_{\nu }R_{\alpha \beta }\right] _{-}^{+}n_{\mu } \\
+\frac{1}{2}\left( \frac{1}{2}h_{\alpha \beta }+n_{\alpha }n_{\beta }\right) 
\overline{\nabla }_{\mu }\left[ R\right] _{-}^{+} \\
-\frac{1}{4}\left[ R\right] _{-}^{+}\left( n_{\alpha }K_{\beta \mu
}+n_{\beta }K_{\alpha \mu }\right) ;
\end{multline}%
\begin{equation}
S_{\alpha }^{\alpha }=0;
\end{equation}%
\begin{multline}
\kappa \left[ S_{\alpha \beta }\right] _{-}^{+}=-\left( \kappa _{1}+\kappa
_{2}\right) \left[ R\right] _{-}^{+}K_{\alpha \beta } \\
+\kappa _{1}n^{\nu }\left[ \nabla _{\nu }R\right] _{-}^{+}+2\kappa
_{2}\delta _{\alpha }^{\rho }\delta _{\beta }^{\sigma }n^{\nu }\left[ \nabla
_{\nu }R_{\rho \sigma }\right] _{-}^{+}.
\end{multline}%
Here, the $S_{\alpha }^{\alpha }$ is the trace of the energy-momentum tensor
of the shell, $S_{\alpha \beta }$, in its distribution form. Also, $\nabla $
and $\overline{\nabla }$ are the covariant derivatives compatible with the
metric $g$\ of the bulks and $h$\ of the shell, respectively. Furthermore, $%
\left[ \Psi \right] _{-}^{+}\equiv \Psi ^{+}-\Psi ^{-}$ denotes a jump in
the function $\Psi $, passing across the thin-shell. Remark that although
all the indices are in Greek, for the shell quantities they only take on the
coordinates on the shell, i.e. $\left\{ t,\theta \right\} $.

For the metric defined in Eq. (2) we calculate the nonzero independent
components of Riemann and Ricci tensors and Ricci scalar as follows 
\end{subequations}
\begin{subequations}
\begin{equation}
R_{trtr}=\frac{1}{2}f^{\prime \prime };
\end{equation}%
\begin{equation}
R_{t\theta t\theta }=\frac{1}{2}ff^{\prime }HH^{\prime };
\end{equation}%
\begin{equation}
R_{r\theta r\theta }=-\frac{H}{2f}\left( 2fH^{\prime \prime }+f^{\prime
}H^{\prime }\right) ;
\end{equation}%
\begin{equation}
R_{tt}=\frac{f}{2H}\left( f^{\prime \prime }H+f^{\prime }H^{\prime }\right) ;
\end{equation}%
\begin{equation}
R_{rr}=-\frac{1}{2fh}\left( f^{\prime \prime }+2fH^{\prime \prime
}+f^{\prime }H^{\prime }\right) ;
\end{equation}%
\begin{equation}
R_{\theta \theta }=-H\left( f^{\prime }H^{\prime }+fH^{\prime \prime
}\right) ;
\end{equation}%
\begin{equation}
R=-\frac{1}{H}\left( f^{\prime \prime }H+2f^{\prime }H^{\prime }+2fH^{\prime
\prime }\right) .
\end{equation}%
Herein, a prime ($^{\prime }$) denotes a total derivative with respect to
the radial coordinate $r$. Now we are at the position to study the JCs in
Eqs. (9).

As discussed in the previous section, the first JCs (Eq. (9a)) require the
continuity of the metric on the shell. Therefore they admit 
\end{subequations}
\begin{equation}
\left\{ 
\begin{array}{c}
f^{+}\left( a\right) =f^{-}\left( a\right) =f_{a} \\ 
H^{+}\left( a\right) =H^{-}\left( a\right) =H_{a}%
\end{array}%
\right. .
\end{equation}%
The second JCs require the first derivatives of the metric functions to be
continuous on the shell, as well; i.e.%
\begin{equation}
\left\{ 
\begin{array}{c}
f^{+\prime }\left( a\right) =f^{-\prime }\left( a\right) =f_{a}^{\prime } \\ 
H^{+\prime }\left( a\right) =H^{-\prime }\left( a\right) =H_{a}^{\prime }%
\end{array}%
\right. \text{ }.
\end{equation}%
The jump in the Ricci scalar is a degree of freedom and considering Eqs.
(10g), (11) and (12) is given by%
\begin{multline}
\left[ R\right] _{-}^{+}=\frac{1}{H_{a}}\times \\
\left[ H_{a}\left( f_{a}^{-\prime \prime }-f_{a}^{+\prime \prime }\right)
+2f\left( H_{a}^{-\prime \prime }-H_{a}^{+\prime \prime }\right) \right] .
\end{multline}%
However, it amounts to%
\begin{equation}
\left[ R\right] _{-}^{+}=0,
\end{equation}%
since the third JCs (Eq. (9c))\ urge%
\begin{equation}
\left\{ 
\begin{array}{c}
f^{+\prime \prime }\left( a\right) =f^{-\prime \prime }\left( a\right)
=f_{a}^{\prime \prime } \\ 
H^{+\prime \prime }\left( a\right) =H^{-\prime \prime }\left( a\right)
=H_{a}^{\prime \prime }%
\end{array}%
\right. .
\end{equation}%
With the results obtained so far, the fourth and fifth JCs (Eqs. (9d) and
(9e))\ are self-satisfied.

For a perfect fluid on a 1+1 hypersurface, the energy-momentum tensor is $%
S_{\alpha }^{\beta }=diag\left( -\sigma ,p\right) $, with $\sigma $ and $p$
being the circumferential energy density and the angular pressure on the
shell, respectively. Hence, the sixth JC (Eq. (9f)) simply implies%
\begin{equation}
\sigma =p.
\end{equation}%
This, of course, is nothing but the static equation of state (EoS) of the
matter on the throat.

Finally, the seventh JCs (Eq. (9g)) amount to the equations%
\begin{multline}
\sigma =\frac{\varsigma }{4\kappa m^{2}}\sqrt{f_{a}}\times \\
\left[ 3\left( f_{a}^{+\prime \prime \prime }-f_{a}^{-\prime \prime \prime
}\right) +\frac{2f_{a}}{H_{a}}\left( H_{a}^{+\prime \prime \prime
}-H_{a}^{-\prime \prime \prime }\right) \right]
\end{multline}%
and%
\begin{multline}
p=-\frac{\varsigma }{4\kappa m^{2}}\sqrt{f_{a}}\times \\
\left[ \left( f_{a}^{+\prime \prime \prime }-f_{a}^{-\prime \prime \prime
}\right) +\frac{6f_{a}}{H_{a}}\left( H_{a}^{+\prime \prime \prime
}-H_{a}^{-\prime \prime \prime }\right) \right] .
\end{multline}%
However, together with Eq. (16), these two equations result in a strong
condition as follows%
\begin{equation}
f_{a}^{+\prime \prime \prime }-f_{a}^{-\prime \prime \prime }-\frac{2f_{a}}{%
H_{a}}\left( H_{a}^{+\prime \prime \prime }-H_{a}^{-\prime \prime \prime
}\right) =0.
\end{equation}%
For both choices $H\left( r\right) =r$ and $H\left( r\right) =1$, the
condition above leads to%
\begin{equation}
f^{+\prime \prime \prime }\left( a\right) =f^{-\prime \prime \prime }\left(
a\right) =f_{a}^{\prime \prime \prime }.
\end{equation}%
This in turn makes both $\sigma $ and $p$ null, i.e.%
\begin{equation}
\sigma =p=0
\end{equation}%
according to Eqs. (17) and (18). Hence, not only for the solutions we have
reviewed in section $II$, but also for any other solutions in the form of
Eq. (2), with an $H\left( r\right) $ function less than (at least) cubic in $%
r$, the existence of the TS is jeopardized as if it had never existed. This
also could be confirmed taking into account the Eqs. (11), (12) and (15),
explicitly. These equations amount to%
\begin{equation}
c_{2}^{+}=c_{2}^{-},
\end{equation}%
\begin{equation}
c_{1}^{+}=c_{1}^{-},
\end{equation}%
and%
\begin{equation}
c_{0}^{+}=c_{0}^{-},
\end{equation}%
which imply that the inner and outer regions are in fact one spacetime, with
no TS, whatsoever.

\subsubsection{TSW construction with RSV-JCs}

In the literature of TSWs, the two spacetimes on the sides of the throat are
traditionally considered to be exact copies of each other. However, it is
shown that this mirror symmetry can be broken by assigning different
spacetimes to $\left( \Sigma ,g\right) ^{\pm }$, and develop ATSWs \cite%
{Forghani1,Forghani3}. In this section we consider this rather generalized
type of TSWs.

The JCs in Eqs. (9) can also be applied to TSWs with some slight
modifications to deal with the discontinuity in the normal vector. The
proper JCs are therefore 
\begin{subequations}
\begin{equation}
\left[ g_{\alpha \beta }\right] _{-}^{+}=0;
\end{equation}%
\begin{equation}
\left[ K_{\alpha \beta }\right] _{-}^{+}=0;
\end{equation}%
\begin{multline}
\left[ R_{\alpha \beta \mu \nu }\right] _{-}^{+}=\frac{\left[ R\right]
_{-}^{+}}{4}\times \\
\left( n_{\alpha }n_{\mu }h_{\beta \nu }-n_{\beta }n_{\mu }h_{\alpha \nu
}-n_{\alpha }n_{\nu }h_{\beta \mu }-n_{\beta }n_{\nu }h_{\alpha \mu }\right)
;
\end{multline}%
\begin{equation}
\left[ R_{\alpha \beta }\right] _{-}^{+}=\frac{\left[ R\right] _{-}^{+}}{2}%
\left( \frac{1}{2}h_{\alpha \beta }+n_{\alpha }n_{\beta }\right) ;
\end{equation}%
\begin{multline}
\left[ \nabla _{\mu }R_{\alpha \beta }\right] _{-}^{+}=\left[ n^{\nu }\nabla
_{\nu }R_{\alpha \beta }\right] _{-}^{+}n_{\mu } \\
+\frac{1}{2}\left( \frac{1}{2}h_{\alpha \beta }+n_{\alpha }n_{\beta }\right) 
\overline{\nabla }_{\mu }\left[ R\right] _{-}^{+} \\
-\frac{1}{4}\left[ R\right] _{-}^{+}\left( n_{\alpha }K_{\beta \mu
}+n_{\beta }K_{\alpha \mu }\right) ;
\end{multline}%
\begin{equation}
S_{\alpha }^{\alpha }=0;
\end{equation}%
\begin{multline}
\kappa \left[ S_{\alpha \beta }\right] _{-}^{+}=-\left( \kappa _{1}+\kappa
_{2}\right) \left[ RK_{\alpha \beta }\right] _{-}^{+} \\
+\kappa _{1}\left[ n^{\nu }\nabla _{\nu }R\right] _{-}^{+}+2\kappa
_{2}\delta _{\alpha }^{\rho }\delta _{\beta }^{\sigma }\left[ n^{\nu }\nabla
_{\nu }R_{\rho \sigma }\right] _{-}^{+}.
\end{multline}%
Imposing the first JCs (Eq. (25a)), analogous to the TS case, necessitates 
\end{subequations}
\begin{equation}
\left\{ 
\begin{array}{c}
f^{+}\left( a\right) =f^{-}\left( a\right) =f_{a} \\ 
H^{+}\left( a\right) =H^{-}\left( a\right) =H_{a}%
\end{array}%
\right. .
\end{equation}%
However, the second JCs (Eq. (25b)) compel a different result as%
\begin{equation}
\left\{ 
\begin{array}{c}
f^{+\prime }\left( a\right) =-f^{-\prime }\left( a\right) \\ 
H^{+\prime }\left( a\right) =-H^{-\prime }\left( a\right)%
\end{array}%
\right. .
\end{equation}%
It is convenient to construct the ATSW with two spacetimes with same $%
H\left( r\right) $ functions on the sides. Hence, we require $H^{+}\left(
r_{+}\right) =H^{-}\left( r_{-}\right) $, which with Eq. (27) admits%
\begin{equation}
H^{+}\left( r_{+}\right) =H^{-}\left( r_{-}\right) =H_{0},
\end{equation}%
where $H_{0}$\ is an arbitrary constant. Accordingly, the second condition
in Eq. (26) is also self-satisfied. With this assumption, exploiting Eq.
(13) for $\left[ R\right] _{-}^{+}$ and the JCs for Riemann tensor in Eq.
(25c), result in%
\begin{equation}
f^{+\prime \prime }\left( a\right) =f^{-\prime \prime }\left( a\right)
=f_{a}^{\prime \prime },
\end{equation}%
and consequently%
\begin{equation}
\left[ R\right] _{-}^{+}=0.
\end{equation}%
While the JCs for the Ricci tensor components and their covariant
derivatives (Eqs. (25d) and (25e)) are automatically satisfied, the JC for
the trace of the energy-momentum tensor of the throat (Eq. (25f)) implies%
\begin{equation}
\sigma =p;
\end{equation}%
similar to the TS case (Eq. (16)). Finally, the last of JCs (Eq. (25g))\
give explicit terms for energy density and tangential pressure\ as%
\begin{equation}
\sigma =\frac{3\varsigma }{4\kappa m^{2}}\sqrt{f_{a}}\left( f_{a}^{+\prime
\prime \prime }+f_{a}^{-\prime \prime \prime }\right)
\end{equation}%
and%
\begin{equation}
p=-\frac{\varsigma }{4\kappa m^{2}}\sqrt{f_{a}}\left( f_{a}^{+\prime \prime
\prime }+f_{a}^{-\prime \prime \prime }\right) ,
\end{equation}%
respectively. However, simultaneous consideration of Eqs. (31), (32) and
(33) suggests%
\begin{equation}
f_{a}^{+\prime \prime \prime }=-f_{a}^{-\prime \prime \prime },
\end{equation}%
which in turn leads to the static EoS%
\begin{equation}
\sigma =p=0.
\end{equation}%
Considering all the results above, imposes conditions on the metric function
coefficients $c_{1}^{\pm }$ and $c_{2}^{\pm }$, as well as the radius of the
ATSW, as follows 
\begin{subequations}
\begin{equation}
c_{2}^{+}=c_{2}^{-},
\end{equation}%
\begin{equation}
c_{1}^{-}=-c_{1}^{+}-2c_{2}^{+}a,
\end{equation}%
and%
\begin{multline}
a=\frac{-c_{1}^{+}\pm \sqrt{c_{1}^{+2}-2c_{2}^{+}\left(
c_{0}^{+}-c_{0}^{-}\right) }}{2c_{2}^{+}} \\
=\frac{-c_{1}^{-}\pm \sqrt{c_{1}^{-2}-2c_{2}^{-}\left(
c_{0}^{-}-c_{0}^{+}\right) }}{2c_{2}^{-}},
\end{multline}%
respectively. Obviously, the radius is real only for $c_{1}^{+2}-2c_{2}^{+}%
\left( c_{0}^{+}-c_{0}^{-}\right) \geq 0$. The TSW radius for the maximally
symmetric case $c_{0}^{+}=c_{0}^{-}$\ amounts to the non-trivial result 
\end{subequations}
\begin{equation}
a=-\frac{c_{1}^{+}}{c_{2}^{+}}=-\frac{c_{1}^{-}}{c_{2}^{-}},
\end{equation}%
which is positive only when $c_{1}^{\pm }$ and $c_{2}^{\pm }$ have different
signs. Since $c_{2}^{+}=c_{2}^{-}$ this alludes $c_{1}^{+}=c_{1}^{-}$, and
the TSW is symmetric. For $H\left( r\right) =1$ and $\lambda =-1$\ this
explicitly becomes%
\begin{equation}
a=-\frac{c_{1}}{4\varsigma m^{2}}.
\end{equation}%
This is a strong condition which dictates on the radius of the TSW. Note
that the signs of the parameters included must eventually be set such that $%
a>0$. Depending on the sign of the quadratic term in the metric function $f$%
, this can be the maximum or the minimum of $f$. Here we emphasize that one
may consider both possibilities $m^{2}>0$\ and\ $m^{2}<0$, for the plus sign
behind the quadratic terms in the CNMG action in Eq. (1) is more of a
convention.

The above results imply, that firstly, the TSW is generally an asymmetric
one, except for the maximally symmetric case where $c_{0}^{+}=c_{0}^{-}$, $%
c_{1}^{+}=c_{1}^{-}$ and $c_{2}^{+}=c_{2}^{-}$, and secondly, depending on
the values of the metric coefficients, the TSW's radius can actually be real
and positive. Providing the tuned up coefficients support the TSW's
existence, it will have null energy density and pressure on its throat,
providing a vacuum condition. Unexpected as it is, now the TSW indeed
satisfies all the energy conditions. This represents a natural wormhole \cite%
{Kim1} with no matter on its throat. The two spacetimes are joined smoothly
and the result is a complete Riemannian manifold with no exotic matter, no
discontinuity or singularity of any sort.

\subsection{DSS Junction Conditions}

In \cite{Deruelle1} DSS have investigated the JCs for the quadratic
Lagrangian density%
\begin{equation}
\mathcal{L}_{\text{DSS}}=\sqrt{-g}\left( R-2\Lambda -4\beta R_{\mu \nu
}R^{\mu \nu }+\alpha R^{2}\right) ,
\end{equation}%
where $\alpha $\ and $\beta $\ are two free parameters and $\Lambda $\
resembles a \textquotedblleft bare\textquotedblright\ cosmological constant.
To do so, they considered the Gaussian-normal coordinates to express the
bulk metrics as 
\begin{equation}
ds^{2}=dy^{2}+h_{ij}dx^{i}dx^{j},
\end{equation}%
in which there exists a thin-shell located at $y=0$, and $h_{ij}$\
represents the metric tensor of the 2-dimensional sub-spacetime. The proper
JCs are found to be%
\begin{equation}
4\left[ -\beta \mathcal{H}_{ij}+\left( \alpha -\beta \right) h_{ij}\mathcal{H%
}\right] _{-}^{+}=S_{ij}
\end{equation}%
where 
\begin{equation}
\mathcal{H}_{ij}\equiv -\frac{1}{2}\frac{\partial ^{3}h_{ij}}{\partial y^{3}}%
,
\end{equation}%
\begin{equation}
\mathcal{H}\equiv h^{ij}\mathcal{H}_{ij},
\end{equation}%
and $S_{ij}$ is the total energy-momentum tensor on the shell.

With a brief comparison between the Lagrangian density of Eq. (39) and the
action of CNMG given in Eq. (1), one finds $\alpha =-\frac{3\varsigma }{%
8m^{2}}$, $\beta =-\frac{\varsigma }{4m^{2}}$, and of course $\Lambda
=\varsigma \lambda m^{2}$. Therefore, the JCs for CNMG can be written as 
\begin{subequations}
\begin{equation}
\frac{\varsigma }{m^{2}}\left[ \mathcal{H}_{i}^{j}-\frac{1}{2}\delta _{i}^{j}%
\mathcal{H}\right] _{-}^{+}=S_{i}^{j}.
\end{equation}%
However, note that prior to checking for the JCs in Eq. (44a) one must check
for the continuity of the metric, and its first and second derivatives with
respect to the normal coordinate $y$\ at the shell's position; i.e. we must
have 
\begin{equation}
\left[ h_{ij}\right] _{-}^{+}=0,
\end{equation}%
\begin{equation}
\left[ \frac{\partial h_{ij}}{\partial y}\right] _{-}^{+}=0,
\end{equation}%
and%
\begin{equation}
\left[ \frac{\partial ^{2}h_{ij}}{\partial y^{2}}\right] _{-}^{+}=0.
\end{equation}%
The JCs in Eq. (44a) can explicitly be determined as 
\end{subequations}
\begin{equation}
\sigma =p=\frac{\varsigma }{2m^{2}}\left[ \mathcal{H}_{\theta }^{\theta }-%
\mathcal{H}_{t}^{t}\right] _{-}^{+}.
\end{equation}%
Comparing the bulk metrics in Eqs. (2) and (40) admits 
\begin{equation}
dy^{2}=\frac{1}{f}dr^{2}
\end{equation}%
and so%
\begin{equation}
\frac{dr}{dy}=\sqrt{f}.
\end{equation}%
This also casts the metric of the TS(W) as%
\begin{equation}
h_{ij}dx^{i}dx^{j}=-fdt^{2}+H^{2}d\theta ^{2},
\end{equation}%
In case of a TS, the continuity of the metric and its first and second
derivatives with respect to the normal coordinate $y$ (JCs in Eqs. (44b-d)),
necessitate%
\begin{equation}
\left\{ 
\begin{array}{c}
f^{+}\left( a\right) =f^{-}\left( a\right) =f_{a} \\ 
H^{+}\left( a\right) =H^{-}\left( a\right) =H_{a}%
\end{array}%
\right. ,
\end{equation}%
\begin{equation}
\left\{ 
\begin{array}{c}
f^{+\prime }\left( a\right) =f^{-\prime }\left( a\right) =f_{a}^{\prime } \\ 
H^{+\prime }\left( a\right) =H^{-\prime }\left( a\right) =H_{a}^{\prime }%
\end{array}%
\right. ,
\end{equation}%
and%
\begin{equation}
\left\{ 
\begin{array}{c}
f^{+\prime \prime }\left( a\right) =f^{-\prime \prime }\left( a\right)
=f_{a}^{\prime \prime } \\ 
H^{+\prime \prime }\left( a\right) =H^{-\prime \prime }\left( a\right)
=H_{a}^{\prime \prime }%
\end{array}%
\right. .
\end{equation}%
Accordingly, one obtains%
\begin{equation}
c_{2}^{+}=c_{2}^{-},
\end{equation}%
\begin{equation}
c_{1}^{+}=c_{1}^{-},
\end{equation}%
and%
\begin{equation}
c_{0}^{+}=c_{0}^{-}.
\end{equation}%
These amount to%
\begin{equation}
\sigma =p=-\frac{\varsigma }{4m^{2}}\frac{f_{a}^{3/2}}{H_{a}}\left(
H_{a}^{+\prime \prime \prime }-H_{a}^{-\prime \prime \prime }\right) ,
\end{equation}%
in consideration of the JCs in Eq. (44a) and the explicit form%
\begin{equation*}
\mathcal{H}_{ij}=-\frac{1}{2}\left( \left( \left( \frac{\partial h_{ij}}{%
\partial r}\right) \sqrt{f}\right) ^{\prime }\sqrt{f}\right) ^{\prime }\sqrt{%
f}.
\end{equation*}%
Here the prime ($^{\prime }$) indicates a total derivative with respect to
the radial coordinate $r$\ and the chain rule is applied. This leads to%
\begin{equation}
\sigma =p=0,
\end{equation}%
for both cases, $H\left( r\right) =r$ and $H\left( r\right) =1$, we have
considered here. Hence, our results using DSS-JCs, analogous to the previous
section, suggest nonexistence of any TS within the framework of CNMG for
special solutions in Eqs. (2) and (3).

For a general ATSW at $r=a$ (where $r_{2}<a<r_{1}$ in case $\varsigma =-1$,
and $r_{1}<a$ in case $\varsigma =1$), the last three JCs give rise to the
exact same results as the previous section's for the metric functions and
their derivatives as%
\begin{equation}
H^{+}\left( r_{+}\right) =H^{-}\left( r_{-}\right) =H_{0},
\end{equation}%
\begin{equation}
f^{+}\left( a\right) =f^{-}\left( a\right) =f_{a}\text{ },
\end{equation}%
\begin{equation}
f^{+\prime }\left( a\right) =-f^{-\prime }\left( a\right) \text{,}
\end{equation}%
and%
\begin{equation}
f^{+\prime \prime }\left( a\right) =f^{-\prime \prime }\left( a\right)
=f_{a}^{\prime \prime },
\end{equation}%
if again the rather physical assumption $H^{+}\left( r_{+}\right)
=H^{-}\left( r_{-}\right) $ holds. Note that, each time a derivative with
respect to the Gaussian normal coordinate $y$\ is taken, the discontinuity
in the normal vector to the throat must be considered, which emerges as the
minus sign in the right-hand-side of Eq. (59). So far, it has been cleared
up that only for the special choice $H(r)=1$ the TSW can be constructed, and
the circumstances and discussions after Eq. (38) in the previous section are
also valid here. Finally, the original JCs in Eq. (44a) impose 
\begin{equation}
\sigma =p=0,
\end{equation}%
which is in full agreement with the previous results.

It appears to us, that using DSS-JCs, the same analysis can be applied to a
wider range of massive quadratic Lagrangian densities and their solutions.
To clear things up, let us consider a massive Lagrangian density of the form%
\begin{equation}
\mathcal{L=}\sqrt{-g}\left( \varsigma R+\frac{1}{m^{2}}\left( R_{\mu \nu
}R^{\mu \nu }+\gamma R^{2}\right) -2\Lambda \right) .
\end{equation}%
Any solution to this Lagrangian density with the form in Eq. (2) and $%
H\left( r\right) =r$ to be used to construct a TSW will suffer from a
discontinuity in the first derivative of the angular component of the TSW
metric. Hence, the natural unsatisfactory behavior of such solutions to the
JCs affects the occurrence of TSW, as if it had never existed.

On the contrary, solutions of the same type with $H\left( r\right) =1$
satisfy all the boundary conditions, but identically lead to%
\begin{equation}
\sigma =p=0.
\end{equation}%
The same analysis, however, is not applicable to RSV-JCs, since for a
general quadratic Lagrangian density such as the one in Eq. (62), the value
of coefficient $\gamma $\ alters the JCs accordingly.

\section{Conclusion}

It has been a long-standing challenge to obtain TSWs with physical, i.e.
non-exotic matter in Einstein's general relativity. This was overcome in the
past in particular models by changing the topology of the shell from
spherical (in 3+1-dimensions) \cite{Mazhari1} and from circular
(2+1-dimensions) forms \cite{Mazhari2,Mazhari3}. Giving up those symmetric
topologies, however, gave rise to different problems in connection with
their stability analysis. In the present study we have shown that the exotic
matter problem is overcome for TSWs in CNMG. In the meantime, the JCs are
modified and they are distinct from those of Einstein's general relativity,
i.e. the DI-JCs. The new JCs are redefined and applied to some static
solutions of CNMG. Our results show no indications of major difference
between the two distinct sets of JCs we have used independently. However,
this is mostly for the quadratic nature of the metric function $f\left(
r\right) $,\ and the specific selection of the gauge function $H\left(
r\right) $. This can be seen the best in the structural differences between
the expressions found for $\sigma $ and $p$\ in Eqs. (32) and (33) using
RSV-JCs, and Eq. (55) using DSS-JCs. More noticeably, the exotic matter
nightmare gets resolved for TSWs in this theory, in the sense that the
energy density and lateral pressure on the shell become zero (better than
negative!), hence no known energy condition is violated anymore.
Nevertheless, these TSWs could only be constructed for the\ gauge selection $%
H\left( r\right) =1$. It was observed that for $H\left( r\right) =r$, which
specifically comprises AdS solutions, no TSWs can be established.\ The
existed TSWs, however, could be symmetric as well as asymmetric. On the
other hand, it was shown that for the specific class of static solutions we
considered here, there cannot be a TS. We leave the profound question of
"how this nonexistence for AdS bulk translates into its CFT correspondence"
for further studies. As a next step, investigating TS and/or TSWs'
constructions under naturally different geometries, such as Lifshitz black
holes \cite{Ayon-Beato1}, is in order. Studying thin-shells with double
layers may also be of interest, as for these ones demand some other JCs \cite%
{Reina1}. As another subject for further studies one may have a look into
extended theories of NMG and solutions therein \cite%
{Gullu1,Gullu2,Ghodsi1,Sinha1,Nam1,Anastasiou1,Setare1,Dalmazi1,Ahmedov4,Louzada1,Camara1,Ghodsi2,Bakas1,Akhavan1,Dereli1,Dereli2}%
.

\bigskip

\bigskip

\end{document}